\documentclass[%
 reprint,
 amsmath,amssymb,
 aip,
]{revtex4-2}

\usepackage{graphicx}
\usepackage{dcolumn}
\usepackage{bm}
\usepackage[dvipsnames]{xcolor}
\usepackage{comment}
\usepackage{gensymb}
\usepackage[utf8]{inputenc}
\usepackage[T1]{fontenc}
\usepackage{mathptmx}
\usepackage{xcolor}   

\newcommand{\siot}{SiO\textsubscript{2}}
\newcommand{\sitnf}{Si\textsubscript{3}N\textsubscript{4}}
\newcommand{\chft}{CHF\textsubscript{3}}
\newcommand{\ot}{O\textsubscript{2}}

\begin{document}

\title{Efficiently-coupled microring circuit for on-chip cavity QED with trapped atoms}

\author{Tzu-Han Chang}
\author{Xinchao Zhou}
\author{Ming Zhu}
\author{Brian M. Fields}
\affiliation{Department of Physics and Astronomy, Purdue University, West Lafayette, IN 47907, USA}
\author{Chen-Lung Hung}
\email{clhung@purdue.edu}
\affiliation{Department of Physics and Astronomy, Purdue University, West Lafayette, IN 47907, USA}
\affiliation{
Purdue Quantum Science and Engineering Institute, Purdue University, West Lafayette, IN 47907, USA}
\affiliation{Birck Nanotechnology Center, Purdue University, West Lafayette, IN 47907, USA}

\date{\today}

\begin{abstract}
We present a complete fabrication study of an efficiently-coupled microring optical circuit tailored for cavity quantum electrodynamics (QED) with trapped atoms. The microring structures are fabricated on a transparent membrane with high in-vacuum fiber edge-coupling efficiency in a broad frequency band. In addition, a bus waveguide pulley coupler realizes critical coupling to the microrings at both of the cesium D-line frequencies, while high coupling efficiency is achieved at the cesium `magic' wavelengths for creating a lattice of two-color evanescent field traps above a microring. The presented platform holds promises for realizing a robust atom-nanophotonics hybrid quantum device.
\end{abstract}

\maketitle

Cold atoms trapped and interfaced with light in photonic optical circuits form exciting hybrid quantum platforms for quantum optics and atomic physics. Strong optical confinement in nanophotonic waveguides or resonators greatly enhances atom-light coupling beyond those achieved in diffraction-limited optics, enabling opportunities in studying light-matter interactions \cite{Thompson2013, Lukin2014Nature, Goban2014NC, nanofiber2014NC,  SuperradiancePRL2015, Circulator2016Science, Laurat2019, LukinPRL2020} and radiative processes \cite{perez2017ultracold,grandi2019hybrid,Ming}. On the circuit-level, nanophotonic engineering offers a variety of tools in modifying the photonic density of states \cite{PhCBook, RMP2018}, as well as controlling photon transport and device optical links \cite{hafezi2013imaging,ozawa2019topological,elshaari2020hybrid,stern2013nanoscale,solano2017super}, thus enriching the complexity of atom-photon interactions and quantum functionality. The indistinguishability and long coherence time of neutral atoms make an atom-nanophotonic hybrid platform inherently scalable, and by itself a strongly coupled many-body system \cite{D.Chang2015, Hung2016, RMP2018}. Recent developments in suspended photonic crystal waveguides and microring resonators \cite{SPYu2014APL, yu2019two, Chang:19, luan2020integration, LukinPRL2020} hold great promises in realizing highly coherent quantum circuits with cold atoms in cavity QED and waveguide QED settings \cite{RMP2018}. 

\begin{figure*}
\centering
\includegraphics[width=0.85\textwidth]{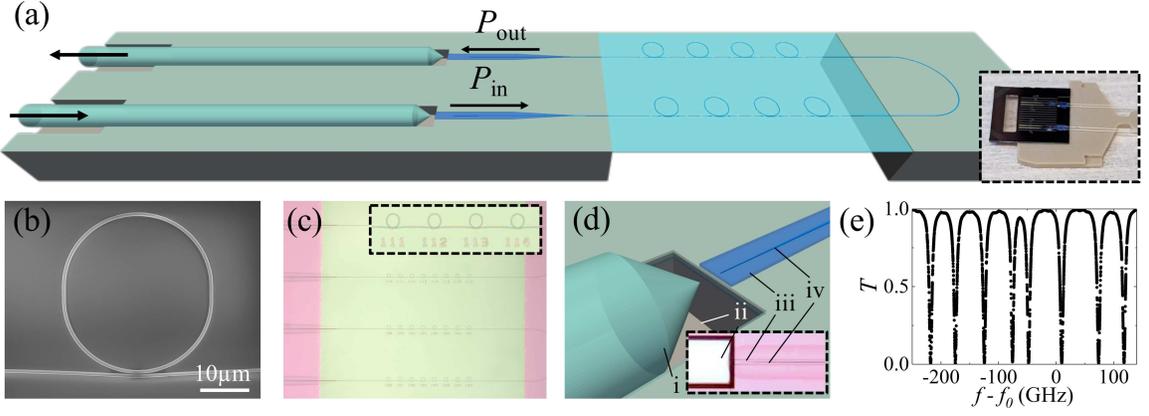}
\caption{(a) Schematics showing the microring resonators (b) on a transparent membrane, the coupling bus waveguide (c), and the top-cladded lens fiber edge-couplers (d) for optical input and output (arrows). The inset shows a fabricated circuit. (b) Scanning electron micrograph of a microring resonator and a bus waveguide in a pulley geometry. Short linear segments in the microring are added to fine-tune the resonator length. (c) Optical micrograph and a zoom-in view (inset) of the microring array. (d) Schematics and micrograph (inset) of the edge-coupler, displaying the lensed fiber (i), the U-shaped fiber groove (ii), and the HSQ top-cladding layer (iii) covering the tapered bus waveguide (iv). (e) Normalized bus waveguide transmission $T=P_\mathrm{out}/P_\mathrm{in}$ versus laser frequency $f$, showing multiple resonances each from a different microring; $f_0 = 335.116$~THz is near the Cs D1-line frequency.
}
\label{fig:sch}
\end{figure*}

Engineering an integrated photonic circuit that fulfills all technical requirements has so-far remained a challenging task. Ideally, the circuit geometry should be compatible with atom cooling and trapping. Nanophotonic waveguides and resonators must be fabricated with high precision, and offer sufficient tunability for alignment with narrow atomic spectral lines. To perform quantum operations with high fidelity, photons should be coupled into and out of a circuit with high efficiency. It is advantageous that a photonic nanostructure could also be utilized to create far off-resonant optical traps to localize cold atoms\cite{Hakuta2004, twocolor2010}. Taking cesium atoms for example, a two-color evanescent field trap formed using the `magic' wavelengths ($\lambda_\mathrm{b}\approx$794~nm and $\lambda_\mathrm{r}\approx$935~nm) 
can create a better trap for coherent quantum operations \cite{twocolor2012, Hung_2013NJP}. As such, all coupling elements to the circuit should work in a broad frequency band. 

\begin{figure}[ht]
\centering
\includegraphics[width=1\columnwidth]{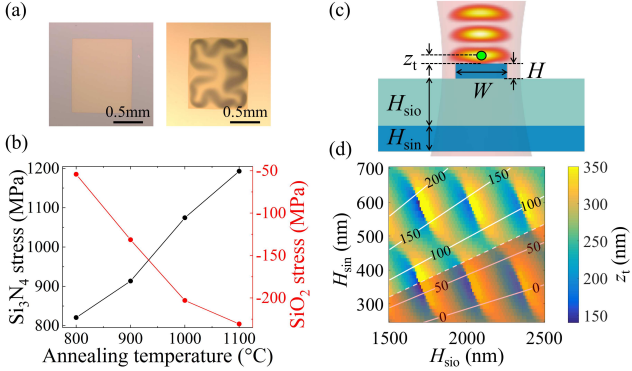}
\caption{(a) Optical micrographs of released membranes with tensile (left, 180 MPa) and compressive (right, -100MPa) resulting stress. (b) Intrinsic stress of LPCVD \sitnf~and \siot~layers measured after post-annealing at various temperatures. (c) Illustration of a lattice of micro traps formed in a top-illuminating optical beam. Position $z_\mathrm{t}$ of the first trap center (green sphere) can be tuned by the layer thickness $(H_\mathrm{SiO},H_\mathrm{SiN})$, as shown in (d). Solid lines indicate constant resulting stress (labeled in MPa) while shaded area marks the unstable region.
}
\label{fig:mem}
\end{figure}

In the letter, we discuss design and full fabrication procedures of an efficiently-coupled microring optical circuit that meets the above key requirements for building a robust hybrid quantum device. An overview of our platform is shown in Fig.~\ref{fig:sch}, where \sitnf~microring resonators are evanescently coupled to a bus waveguide in a pulley geometry for optical input and output [Fig.~\ref{fig:sch}(b-c)]. The microrings are top vacuum-cladded and are fabricated on a transparent \siot-\sitnf~double-layer membrane (Fig.~\ref{fig:sch}(c) and Fig.~\ref{fig:mem}), suspended over a large window (2~mm$\times$8~mm) on a silicon chip. This ensures full optical access for laser cooling and cold atom trapping \cite{Kim2019}. The microring geometry is designed to optimize the cooperativity parameter $C = \frac{3\lambda^3}{4\pi^2}\frac{Q}{V_\mathrm{m}}$ for cavity QED with cesium atoms \cite{Chang:19}, where $\lambda=\lambda_\mathrm{D1} = 894~$nm ($\lambda_\mathrm{D2} = 852$~nm) is the wavelength of Cs D1 (D2) line. The microring radius is $R\approx 15~\mu$m and the waveguide width and height are $(W,H)=(750,380)~$nm, respectively (Fig.~\ref{fig:mem}). A nearly optimal $Q/V_\mathrm{m}$ ratio is achieved, giving $C\approx 10\sim30$ with an intrinsic quality factor $Q\approx 1\sim 3.3\times 10^5$, currently limited by the surface scattering loss\cite{Chang:19}, and a mode volume $V_m \approx 500~\mu$m$^{3}$, evaluated using the normalized transverse-magnetic (TM) mode field amplitude at an atomic position $z_\mathrm{t}\approx100~$nm centered above the microring dielectric surface. We adopt the fundamental TM-mode for its uniform polarization above a microring \cite{Chang:19}.

By lithographically scanning the length of each microring (Fig.~\ref{fig:sch}(b)), their resonances approach the targeted frequency as shown in the transmission spectrum in Fig.~\ref{fig:sch}(e). Precise alignment to the atomic spectral lines can be thermally tuned, for example, by a laser beam globally heating the silicon substrate under vacuum. The tunability is $\sim 0.5$~GHz/mW. The transmission spectrum is measured through lensed fibers coupled to either end of the bus waveguide via an edge-coupler (Fig.~\ref{fig:sch}(d)). Each of the resonances in Fig.~\ref{fig:sch}(e) displays nearly zero transmission $T\approx 0$, achieving the ideal critical coupling condition for probing atom-microring coupling; see Fig.~\ref{atomlight}. 

We begin the circuit fabrication by preparing for a \siot-\sitnf~double-layer membrane stack, deposited on a silicon wafer using low-pressure chemical vapor deposition (LPCVD) processes. For stable membrane release from the silicon substrate, the compressive stress of the \siot~layer ($\sim2~\mu$m thick) should be overcome by the tensile stress of the \sitnf~bottom-layer, giving a thickness-weighted tensile resulting stress \cite{rossi1998realization}. We arrive at a proper stress condition by post-annealing at around $1100^{\circ}$C for the \sitnf~bottom-layer and at $950^{\circ}$C after we deposited the \siot~layer, two hours for each time (Fig.~\ref{fig:mem}(b)). In our results, a post-annealed membrane can be released free from buckling and severe cracking under a tensile stress of $70\sim180$ MPa.

\begin{figure*}[t]
\centering
\includegraphics[width=0.85\textwidth]{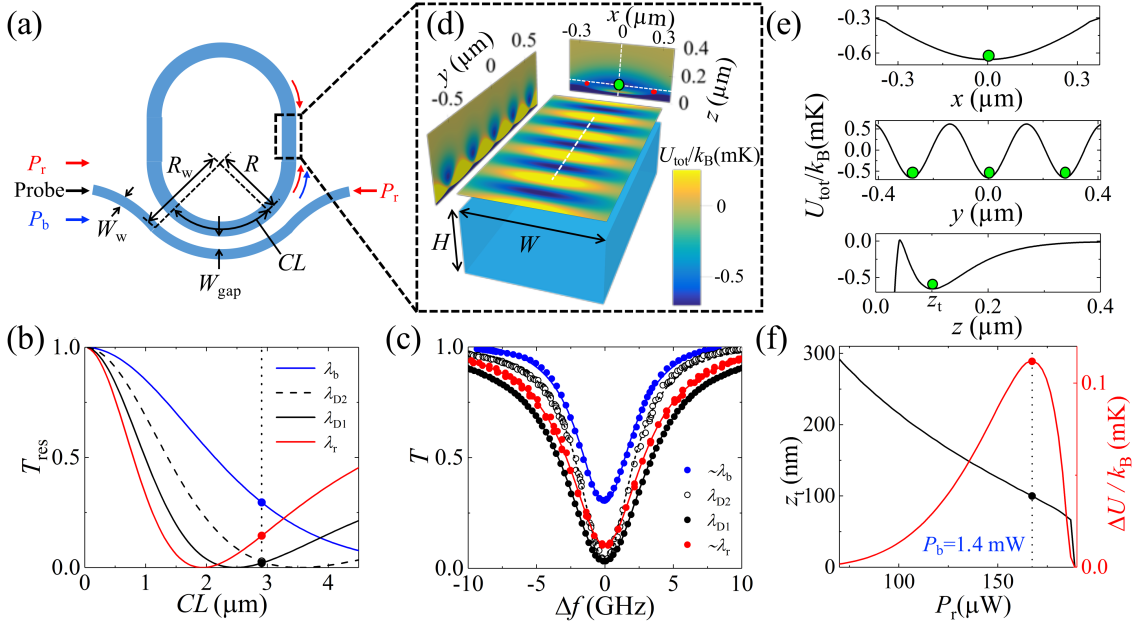}
\caption{(a) Schematics of the pulley-coupler. The coupler gap $W_\mathrm{gap}$, bus waveguide width $W_\mathrm{w}$, and bend radius $R_\mathrm{w}$ are optimized and fixed at $(0.21,0.55,15.86)~\mu$m, respectively, along with the microring waveguide parameters $(W,H,R)=(0.75,0.38,15)~\mu$m. $CL$ is a variable coupling length. Color arrows depict the injected light for the two-color evanescent field trap ($\lambda_\mathrm{b}$: blue, $\lambda_\mathrm{r}$: red) and the atom probe ($\lambda_\mathrm{D1,D2}$: black). (b) Simulated resonant transmission $T_\mathrm{res}$ versus coupling length $CL$ at the indicated wavelengths. Vertical dashed line marks $CL = 2.9~\mu$m for the fabricated device. (c) Measured transmission spectra (color symbols) with laser wavelengths as in (b). The spectra are shifted to display resonances at laser detuning $\Delta f = 0$, showing good agreement with calculation $T(0)\approx T_\mathrm{res}$ (filled circles in (b)). Solid curves are fits to extract $\kappa_c$ and $\kappa_i$. Based on these measurements, a two-color evanescent field trap can be accurately evaluated. (d) Potential cross sections above a linear segment of the microring (dashed box in (a)), calculated using injected power $(P _\mathrm{r},  P_\mathrm{b}) = (0.17 ,1.4)$~mW, respectively. Potential line-cuts along the dashed lines are shown in (e). Green spheres mark the trap centers. Red spheres mark the saddle points where the trap opens. The potential difference between the weakest saddle point and the trap center defines the trap depth \cite{Chang:19}. (f) Variable vertical trap position $z_\mathrm{t}$ (black curve) as indicated in (e) and trap depth $\Delta U$ (red curve), adjusted through the power ratio $P _\mathrm{r}/P_\mathrm{b}$. Dashed line marks the condition at which $z_\mathrm{t} = 100$ nm.
}
\label{fig:pul}
\end{figure*}

Optical reflectance is another crucial factor in determining the membrane thickness, primarily concerning atom trapping. For example, in a top-illuminating optical tweezer trap implemented in Ref.~\cite{Kim2019} or \cite{tweezer2020}, membrane reflection and interference result in a lattice of micro traps formed within a tweezer beam as shown in Fig.~\ref{fig:mem}(c). We scan the \siot/\sitnf~layer thickness to minimize the position $z_\mathrm{t}$ of the first micro trap (formed by an anti-node) in a tweezer potential, while monitoring the resulting stress. An example is shown in Fig.~\ref{fig:mem}(d), calculated for a $\lambda_\mathrm{r} = $935~nm tweezer trap focused by an objective of numerical aperture N.A.$=0.35$. A micro trap at $z_\mathrm{t} \approx 150~$nm forms with layer thickness $(H_\mathrm{SiO},H_\mathrm{SiN}) \approx (1.72,0.55)~\mu$m within the stable membrane regime.

Once the membrane stack is fabricated, an additional LPCVD-grown \sitnf~top-layer is deposited, and the wafer is diced into centimeter-sized chips (Fig.~\ref{fig:sch}(a) inset). Microring arrays and bus waveguides are then fabricated in the top layer using e-beam lithography with multipass writing and an inductively coupled plasma reactive-ion etching (ICP-RIE) process with \chft/\ot~gas chemistry \cite{Ji:17}.

Either end of a bus waveguide is designed to taper down and terminate at a width of 70~nm for edge-coupling with a lensed fiber \cite{6895281} (1~$\mu$m focused beam waist), which is placed inside a U-shaped fiber groove of $\sim 65~\mu$m depth  (Fig.~\ref{fig:sch}(d)).  To achieve high coupling efficiency, a top-cladding structure on each edge-coupler is designed to improve symmetric mode matching with the lensed fiber. The geometry of the top-cladded edge-coupler has been numerically optimized using finite-different-time-domain calculations to achieve $\sim$70\% coupling efficiency at Cs D-lines. The same coupler yields an efficiency $\sim$70\% (60\%) at $\lambda_\mathrm{r}\approx$935~nm ($\lambda_\mathrm{b}\approx$794~nm).

To fabricate the top-cladding structure on each edge-coupler defined in the first lithography step, we cover the tapered bus waveguides with $\sim$1~$\mu$m-thick hydrogen silsesquioxane (HSQ) resist as the top-cladding material. Using second e-beam lithography, we define the top-cladding structures to inversely taper down along the bus waveguides  (Fig.~\ref{fig:sch}(d)) so to keep the microrings top vacuum-cladded. Fiber grooves and edge-coupler facets are defined in a subsequent photolithography step. HSQ and \siot~are then etched away in the ICP-RIE, followed by the Bosch process etching to create fiber U-grooves in the silicon.

Following the fabrication of photonic structures at the front side of the chip, the membrane is then released from the silicon substrate. To do this, a window at the backside of the chip is first defined using photolithography, while the front-side is protected with a thick layer of spin-coated photoresist (PMMA 950 coating with Surpass 3000 adhesion promoter). Materials at the backside are etched away in the ICP-RIE and the Bosch process until leftover silicon is only $\sim15~\mu$m thick. To gently release the membrane, we perform wet-etching in a 12\% aqueous TMAH solution at 65 \degree C, followed by DI water rinsing and cleaning with PRS-2000 stripper and Nanostrip to expose the front-side photonic structures. Finally, a thin alumina layer ($\sim 5~$nm) is deposited using atomic layer deposition to protect the microrings against cesium corrosion during experiments \cite{woetzel2013lifetime}.

Our fabrication procedure yields nearly 100\% success rate on membrane release. The released membrane is optically flat and has a root-mean-squared surface roughness of $1.4$~nm. Additional chemical mechanical polishing step \cite{Ji:17} can be applied to the \siot-\sitnf~double-layer stack following LPCVD. We have measured surface roughness down below $0.5$~nm, which is expected to improve the microring quality factor to $Q>10^6$ using similar fabrication procedures \cite{Chang:19}.

In the final step, optical lensed fibers are introduced to the fiber grooves. The alignment tolerance is $\sim \pm 0.3~\mu$m for 1-dB excess loss. Fine adjustment in the U-groove is required prior to epoxy fixture. We note that misalignment can occur under vacuum when bulk epoxy outgases and shrinks. Therefore, only a thin layer of low viscosity UV epoxy (OG198-54) is applied for alignment fixture. We achieve $\sim$50\% ($\sim$3dB loss) coupling efficiency per facet, which persists under vacuum pressure below $10^{-6}$ Torr. The fibers are guided out of a vacuum chamber without noticeable loss via a teflon feedthrough mounted on a Swagelok fitting \cite{abraham1998teflon}.

We now discuss the design of pulley couplers (Fig.~\ref{fig:pul}(a)), which allows us to separately optimize the bus waveguide parameters and the coupling length $CL$ for efficient microring coupling over a wide frequency band. We perform a finite element method (FEM) analysis to calculate the microring coupling rate \cite{Ali2010}
$\kappa_c= |S\times \mathrm{sinc}\left [(n_\mathrm{w}R_\mathrm{w} -nR)\frac{CL}{R\lambda}\right ]\frac{CL}{R}|^2$,
where $\mathrm{sinc}(x) = \sin(\pi x)/\pi x$ is the normalized sinc function, $\lambda$ is the coupling wavelength ($\omega$ is the angular frequency),  $S=\frac{i\omega\epsilon_{0}}{4}\int(\epsilon_\mathrm{w}(r,z)-1)\tilde{\mathbf{E}}_{w}\cdot\tilde{\mathbf{E}}^*rdrdz$, and the integration runs over the cross-section of the bus waveguide with $\epsilon_\mathrm{w}(r,z)$ being its dielectric function; $\left(\tilde{\mathbf{E}}_\mathrm{(w)}, n_\mathrm{(w)}, R_\mathrm{(w)}\right)$ correspond to the normalized resonator (bus waveguide)
mode field, the effective refractive index, and the bend radius, respectively. By comparing $\kappa_c$ with the microring intrinsic decay rate $\kappa_i$, also evaluated using a FEM analysis \cite{Chang:19}, we can optimize the bus waveguide-microring coupling efficiency numerically.

Figure \ref{fig:pul}(b) shows the expected resonant transmission $T_\mathrm{res} = |\frac{\kappa_i - \kappa_c}{\kappa_i + \kappa_c}|^2$ as a function of the coupling length $CL$. The better the coupling efficiency, the lower the transmission. Approaching the critical coupling condition ($\kappa_c=\kappa_i$), all of the resonant input photons can be drawn into the microring, hence, resulting in zero transmission $T_\mathrm{res}=0$. In this calculation, the pulley coupler geometry is chosen to improve the overlap of the critical coupling regions ($T_\mathrm{res}\approx 0$) of all four relevant wavelengths. For the fabricated devices, we have selected $CL= 2.9~\mu$m to approach critical coupling for Cs D-lines at $\lambda_\mathrm{D1}$ and $\lambda_\mathrm{D2}$, respectively, while maintaining sufficient coupling efficiency near $\lambda_\mathrm{b}$ and $\lambda_\mathrm{r}$ magic wavelengths for two-color evanescent field traps \cite{Chang:19}. We anticipate $T_\mathrm{res} = (0.02,0.03,0.14,0.29)$ for test wavelengths $\lambda = (894, 852, 932, 795 )~$nm, respectively, which are in very good agreement with the measurement $T(0) \approx (0.03,0.04,0.10,0.31)$ as shown in Fig.~\ref{fig:pul} (c).

The agreement between the bus waveguide transmission measurement and full simulation results illustrate the fabrication precision of our microring optical circuit. Meanwhile, the absence of resonance splitting in the transmission data (Fig.~\ref{fig:pul} (c)) suggests that there is negligible mode-mixing caused by coherent back-scattering in the microring resonator \cite{Chang:19, Painter07}. Therefore, the resonant TM modes preserve the traveling-wave characteristics of a whispering-gallery mode (WGM) \cite{Chang:19}. 

Combining our measurement and FEM simulation results, we can now estimate the actual power required to create a stable two-color evanescent field trap. One sample scheme is illustrated in Fig.~\ref{fig:pul} (a). A resonant TM mode near the wavelength $\lambda_\mathrm{b}$ is excited from the bus waveguide to create a smooth and short-range repulsive optical potential, preventing atoms from crashing onto the microring surface. Additional two phase-coherent, counter-propagating TM modes near the wavelength $\lambda_\mathrm{r}$ are excited from either end of the bus waveguide to create an attractive optical lattice-like potential, localizing atoms tightly above the microring.  We note that, given the finite free spectral range of the microring, to achieve a satisfactory alignment of the microring resonances near the Cs magic wavelengths while maintaining the exact alignment of a resonator mode to the Cs D-line (through thermal tuning) requires further lithographic-tuning of the microring geometrical parameters near the reported values. Fig.~\ref{fig:pul} (d-e) plot the total ground state trap potential $U_\mathrm{tot}$, including the direct summation of repulsive and attractive potentials calculated using the electric field profiles obtained from the FEM analyses and the build-up intensity in the microring with a total input power of $P_\mathrm{b} +2P_\mathrm{r} \approx 1.8~$mW. We have also added in $U_\mathrm{tot}$ an approximate Casimir-Polder attractive potential $U_\mathrm{cp}= - C_4/z(z^3+\lambdabar)$, where $C_4=h \times 267~$Hz$\cdot \mu$m$^4$ is the Cs-\sitnf~ surface interaction coefficient and $\lambdabar = 136~$nm is an effective wavelength \cite{stern_simulations_2011}. 

The two-color evanescent field trap is robust and tunable. The vertical trap location $z_\mathrm{t}$ and the trap depth $\Delta U$ can be finely controlled by the power of injected light $P_{\mathrm{r(b)}}$ (Fig.~\ref{fig:pul}(f)). At an optimal power ratio shown in Fig.~\ref{fig:pul}(e), the trap can be tuned to $z_\mathrm{t}\approx 100~$nm with $\Delta U \approx k_\mathrm{B} \times 120~\mu$K, much deeper than the thermal energy $\lesssim k_\mathrm{B} \times 10~\mu$K of polarization-gradient cooled cesium atoms, where $k_\mathrm{B}$ is the Boltzmann constant.

\begin{figure}[t]
\centering
\includegraphics[width=1\columnwidth]{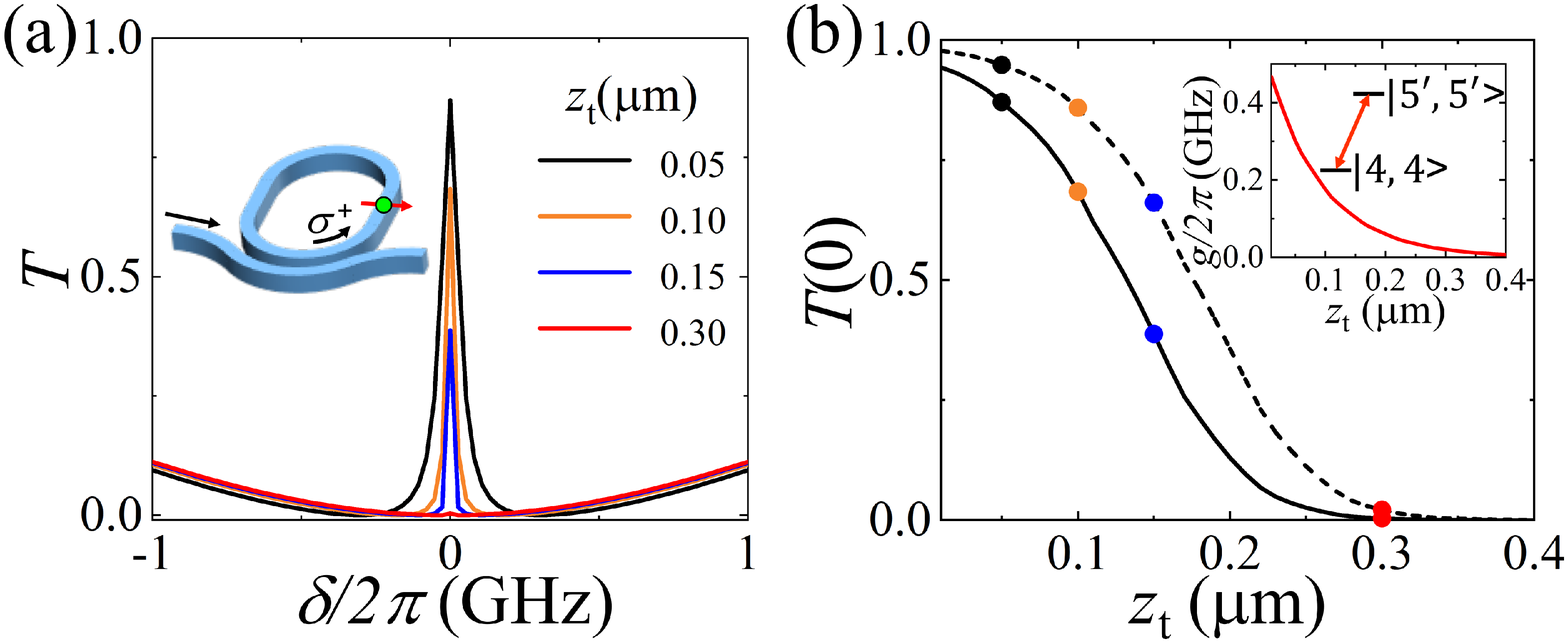}
\caption{(a) Transmission $T$ with a spin-polarized atom trapped at the indicated positions $z_\mathrm{t}$, calculated using $\kappa_c = \kappa_i = 2\pi \times 2.8~$GHz extracted from Fig.~\ref{fig:pul}(c) and the atom-photon coupling strength $g$ for $\sigma^+$-transition driven by the CCW WGM as shown in the inset of (b). The polarization axis (red arrow) of the trapped atom (green sphere), and the CCW WGM (curved arrow) is depicted in the inset. (b) Resonant transmission $T(0)$ versus $z_\mathrm{t}$ (solid curve) and the case for $\kappa_c=\kappa_i= 2\pi \times 1.0~$GHz (dashed curve, for $Q\approx3.3\times 10^5$). Color symbols mark the trap locations as in (a).}\label{atomlight}
\end{figure}

By realizing atom trapping and a critically-coupled microring resonator aligned to an atomic resonance, it is possible to probe the interaction between single atoms and resonator photons with high sensitivity. As a simple example, in Fig.~\ref{atomlight} we plot a weakly-driven, steady-state bus waveguide transmission \cite{aoki2006observation,Ming}, $
    T(\delta) = \left|\frac{g^2 + (i\delta + \frac{\Gamma}{2})(i \delta + \frac{\kappa_i-\kappa_c}{2})}{g^2 + (i\delta + \frac{\Gamma}{2})(i \delta + \frac{\kappa}{2})}\right|^2$,
where $\delta$ is the laser detuning from the atomic resonance (D2 line), $g = \sqrt{\frac{3\lambda_\mathrm{D2}^3\omega\Gamma}{16\pi^2V_m}}$ is the position-dependent atom-photon coupling strength (Fig.~\ref{atomlight}(b) inset),
$\Gamma=2\pi\times 5.2~$MHz is the atomic decay rate, and the resonator decay rate $\kappa=\kappa_\mathrm{i}+\kappa_\mathrm{c}=2\pi \times 5.6~$GHz is extracted from the measurement. A trapped atom is assumed to be initially polarized in the ground state $6S_{1/2}|F=4,m_F=4\rangle$ and is excited to $6P_{3/2} |F'=5,m_{F'}=5\rangle$ by a circularly polarized counter-clockwise (CCW) circulating WGM as shown in Fig.~\ref{atomlight}(a) inset. Significant increase in the bus waveguide transmission $T(0)\approx 1-\kappa\Gamma/2g^2$ can be observed near the atomic resonance when $g^2 > \kappa\Gamma$. This transparency window results from the destructive interference between atom-WGM photon dressed states, similar to the electromagnetically-induced transparency effect \cite{Arno2013PRL, VIT2011Science, EIT2010Nature}. At the desired trap location $z_\mathrm{t}\approx 100~$nm, we expect $g\approx 2\pi \times 176~$MHz and $T(0)\approx 0.68$ ($T(0)\approx 0.86$ for currently best available $\kappa/2\pi\approx2.0~$GHz), which greatly contrasts $T(0)\approx 0$ of an empty microring (Fig.~\ref{atomlight}(b)). The large variation of bus waveguide transmission could thus inform us the presence of single atom and the strength of atom-photon coupling with high sensitivity \cite{aoki2006observation, Shomroni903}. Lastly, we note that the WGM circular polarization results from strong confinement in the microring nanowaveguide; the polarization is locked to the direction of the WGM circulation \cite{VanMechelen:16}. Creating a directional coupling with spin-polarized atoms can give rise to applications, for example, in chiral quantum optics \cite{nanofiber2014NC,Circulator2016Science, Lodahl2017}.

We demonstrate a microring optical circuit that permits precision understanding of fabrication performance and analyses of the optical modes. Our circuit is efficiently coupled, scalable, and can simultaneously accommodate large number of atoms trapped in an array of surface micro traps. With near-term improvement on the membrane surface quality\cite{Chang:19} and reduction of other surface scattering sources\cite{Ji:17,Porkolab:14}, one expects more than ten-fold increase in the quality factor $Q>4\times10^6$ to achieve large cooperativity $C>250$ when the microring is critically coupled. Coherent quantum operations with a single atom or in a hybrid lattice formed by atoms and photons can be realized following introduction of cold atoms to the microring to form a hybrid quantum circuit with strong atom-photon interactions. In the latter case, the hybrid lattice can form a strongly coupled many-body system with WGM photon-mediated interaction among all trapped atoms \cite{D.Chang2015}. Non-uniform or pair-wise tunable interactions can be achieved in this resonator by using local addressing, via top-projected optical tweezers, or a multi-frequency pumping scheme as detailed in Ref.~\cite{Hung2016}. Dynamics of the strongly-coupled hybrid lattice may be probed via the efficiently coupled bus waveguide, or by using single atom-resolved fluorescence imaging \cite{Kim2019,Meng2020}.

Funding is provided by the AFOSR YIP (Grant NO. FA9550-17-1-0298) and the ONR (Grant NO. N00014-17-1-2289). X. Zhou and M. Zhu acknowledge support from the Rolf Scharenberg Graduate Fellowship. The data that support the findings of this study are available from the corresponding author upon reasonable request.

\nocite{*}

\bibliography{aipsamp}

\end{document}